\documentclass{article}

\usepackage{amssymb}
\usepackage{amsmath}

\usepackage{latexsym}
\usepackage{url}

\usepackage{lineno}
\usepackage{eurosym}
\usepackage[ruled,vlined]{algorithm2e}
\usepackage{subfig}
\usepackage{authblk}
\usepackage{enumitem}

\usepackage{graphicx}
\usepackage{hyperref}

\newcommand{\C}{\mathcal{C}}
 
\newcommand{\F}{\mathcal{F}}

\newcommand{\mathsc}[1]{{\normalfont\textsc{#1}}}

\newcommand{\partitions}{p}
\newcommand{\partitionsPerDim}{m}
\newcommand{\cores}{c}

\newcommand{\sky}{\mathsc{Sky}}
\newcommand{\nd}{\mathsc{ND}}
\newcommand{\po}{\mathsc{PO}}

\newcommand{\noseq}{\mathsc{NoSeq}\xspace}

\newcommand{\grid}{\mathsc{Grid}\xspace}
\newcommand{\angular}{\mathsc{Angular}\xspace}
\newcommand{\sliced}{\mathsc{Sliced}\xspace}
\newcommand{\slicedrep}{\mathsc{Sliced+}\xspace}
\newcommand{\angularrep}{\mathsc{Angular+}\xspace}

\newcommand{\dominates}{\prec}

\newcommand{\fdominates}{\dominates_{\F}}

\newcommand{\dimensions}{d}
\newcommand{\monotoneFunctions}{\mathtt{MF}}

\newcommand{\positivereals}{\mathbb{R^+}}

\newcommand{\dominanceRegion}{\mathsc{DR}}

\newcommand{\size}{N}

\newcommand{\mycomment}[1]{}

\newtheorem{theorem}{Theorem}

\newtheorem{definition}{Definition}
\newtheorem{example}[theorem]{Example}

\newcommand{\logSep}{\,.\,\,}

\def\codeforeach{\mbox{\upshape\textbf{for each}}}
\def\codedo{\mbox{\upshape\textbf{do}}}

\def\codereturn{\mbox{\upshape\textbf{return}}}

\def\codeparallel{\mbox{\upshape\textbf{parallel}}}

\newcommand{\sveonef}{\texttt{SVE1F}\xspace}

\newcommand{\popi}{\texttt{POPI2}\xspace}

\providecommand{\keywords}[1]{\textbf{\textit{Keywords---}} #1}

\begin{document}

\title{Partitioning Strategies for Parallel Computation of Flexible Skylines}

\author{Emilio De Lorenzis and Davide Martinenghi}

\urldef{\autone}\url{emilio.delorenzis@mail.polimi.it}
\urldef{\auttwo}\url{davide.martinenghi@polimi.it}

\affil{Politecnico di Milano, DEIB\\Piazza Leonardo 32, 20133 Milan, Italy.\\email:
\autone, 
\auttwo
}
 
\date{}

\maketitle

While classical skyline queries identify interesting data within large datasets, flexible skylines introduce preferences through constraints on attribute weights, and further reduce the data returned. However, computing these queries can be time-consuming for large datasets. We propose and implement a parallel computation scheme consisting of a parallel phase followed by a sequential phase, and apply it to flexible skylines. We assess the additional effect of an initial filtering phase to reduce dataset size before parallel processing, and the elimination of the sequential part (the most time-consuming) altogether.
All our experiments are executed in the PySpark framework for a number of different datasets of varying sizes and dimensions.


\pagestyle{plain}

\section{Introduction}
\label{intro}

With the growth of big data, efficiently finding interesting data within large datasets has become essential. Skyline queries are a method to select a subset of data by returning tuples that are not dominated by any other tuple. A tuple $t$ dominates another tuple $s$ if $t$ is not worse than $s$ in any attribute and is strictly better in at least one. Top-$k$ queries, on the other hand, reduce multi-objective problems to single-objective ones using a scoring function that incorporates parameters like weights to reflect user preferences for different attributes. While skylines provide a global view of potentially interesting data, they do not consider user preferences and may return too many tuples, making it difficult for users to make decisions.

To address these issues, \emph{Flexible Skylines} combine the concepts of skyline and top-$k$ queries by applying constraints to attributes in order to specify preferences and give different importance to each. The concept of $\F$-dominance is the key idea to extend dominance: a tuple $t$ $\F$-dominates another tuple $s$ if $t$ is always better than or equal to $s$ according to all scoring functions in $\F$. Flexible skylines identify a subset of the skyline and exist in two flavors: the non-dominated flexible skyline ($\nd$), which returns the subset of non-$\F$-dominated tuples, and the potentially optimal flexible skyline ($\po$), which returns a subset of $\nd$ representing all tuples that are top-1 with respect to a scoring function in $\F$. Although, as mentioned, the cardinality of interesting tuples is typically smaller than in the case of skylines, flexible skylines are still computationally intensive operators.

In order to tackle the unmanageability of centralized algorithms in the face of very large datasets, we propose a parallel scheme that partitions the data and then treats every partition in parallel. Each partition processes a part of the dataset and returns a so-called ``local'' result, which is then merged in a sequential phase to find the ``global'' result.

Other works have attempted the adoption of partitioning strategies for computing skylines, but this is the first attempt for flexible skylines, which have inherent difficulties of their own.
We also assess the effect of an initial filtering phase to decrease the load on the parallel part, as well as methods to eliminate the sequential phase entirely.
The algorithms are implemented in PySpark, a parallel environment based on Spark, and tested on virtual machines on a datacloud with up to 30 cores.

\section{Background}
\label{sec:prelim}

We focus on on numeric attributes in $\positivereals$ and generally refer to a schema $S$ of attributes $\{A_1,\ldots, A_\dimensions\}$ over such a domain. We use the notion of tuple and relation over $S$ in accordance with the standard relational model, so that $t[A_i]$ is the value of tuple $t$ over attribute $A_i$.

The \emph{skyline}~\cite{DBLP:conf/icde/BorzsonyiKS01} of a relation $r$ is the set of non-dominated tuples in $r$, where we say that $t$ \emph{dominates} $s$, denoted $t\dominates s$, if, for every attribute $A\in S$, $t[A]\leq s[A]$ holds and there exists an attribute $A' \in S$ such that $t[A']<s[A']$ holds:
\begin{equation}\label{eq:skyline}
\sky(r)=\{t \in r \mid \nexists s\in r \logSep s\dominates t\}.
\end{equation}
The score of $t$ through \emph{scoring function} $f$ is the value $f(t[A_1],\ldots, t[A_\dimensions])\in \positivereals$,
also indicated $f(t)$.
We conventionally assume, for both scores and attribute values, that smaller values are preferable.

The skyline can be equivalently defined as the set of tuples that are top-$1$ results for at least one monotone scoring function.
\begin{equation}\label{eq:skyline-alt}
\sky(r) = \{t \mid t\in r \land \exists f\in\monotoneFunctions
\logSep \forall s\in r\logSep s\neq t \implies f(s) > f(t)\},
\end{equation}
where $\monotoneFunctions$ indicates the set of all monotone functions.

The notion of \emph{flexible skyline} was introduced in~\cite{DBLP:journals/pvldb/CiacciaM17} as a generalization of 
Equations~\eqref{eq:skyline} and~\eqref{eq:skyline-alt}:
\begin{definition}\label{def:flexible-skyline}
For a set $\F$ of monotone scoring functions, $t$ \emph{$\F$-dominates} $s$, denoted $t\fdominates s$, iff, $\forall f\in\F\logSep f(t)\leq f(s)$ and $\exists f\in\F\logSep f(t)< f(s)$.
The \emph{non-dominated flexible skyline} $\nd(r;\F)$ 
 is the set $\{t \in r \mid \nexists s\in r \logSep s\fdominates t\}$.
The \emph{potentially optimal flexible skyline} $\po(r;\F)$ 
 is the set 
$\{t\in r \mid \exists f\in\F
 \logSep \forall s\in r\logSep s\neq t \implies f(s) > f(t)\}$.
\end{definition}

We observe that $\nd$ generalizes skylines as in Equation~\eqref{eq:skyline} (with $\fdominates$ instead of $\dominates$), while $\po$ generalizes Equation~\eqref{eq:skyline-alt} (with $\F$ instead of $\monotoneFunctions$).
The \emph{$\F$-dominance region} $\dominanceRegion(t;\F)$ is 
the set of all points $\F$-dominated by $t$.
As a general property, $\po(r;\F)\subseteq\nd(r;\F)\subseteq \sky(r)$.

\begin{figure}
\centering
\hspace*{-0.20\textwidth}
\subfloat[][{Dataset in tabular form.\label{fig:example-table}}]
{ 
\scalebox{0.8}{\vbox to 0.45\textwidth{\vfil
   		\begin{tabular}{r|c|c|}
\multicolumn{1}{c}{$ $}&\multicolumn{1}{c}{$ts$}	&	\multicolumn{1}{c}{$ch$}\\
\cline{2-3}
\texttt{\textit{a}} & 0.30& 0.80\\
\texttt{\textit{b}} & 0.55& 0.45\\
\texttt{\textit{c}} & 0.70& 0.30\\
\texttt{\textit{d}} & 0.40& 0.90\\
\texttt{\textit{e}} & 0.60& 0.20\\
\texttt{\textit{f}} & 0.60& 0.90\\
\texttt{\textit{g}} & 0.90& 0.15\\
\texttt{\textit{h}} & 0.50& 0.70\\
\texttt{\textit{i}} & 0.80& 0.10\\
\cline{2-3}
   		\end{tabular}
   		\vfil
 		}
 		}
 		}%
\hspace*{-0.20\textwidth}
\subfloat[][{2D depiction of the locations.}]
{\includegraphics[width=0.49\textwidth]{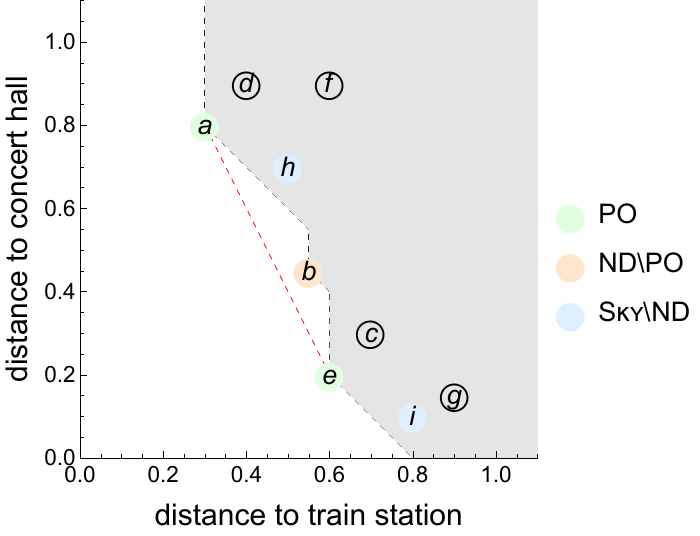}\label{fig:example}}%
 	\caption{A set $r$ of locations and their distances to given points of interest.}
 	\label{fig:sorted-lists}
 \end{figure}

\begin{example}
Consider the dataset $r$ in Figure~\ref{fig:example-table}, showing locations with their distance from points of interest.
The non-dominated options in $r$ are \texttt{a}, \texttt{b}, \texttt{h}, \texttt{e}, \texttt{i}: this is $\sky(r)$.
Consider now the set $\F$ to be the scoring functions of the form $f(x,y)=w_1 x+ w_2 y$, where $w_1>w_2$, i.e., linearly combining distances, with more importance given to the train station.
Under $\F$, \texttt{h} and \texttt{i} are $\F$-dominated by \texttt{a} and \texttt{e}, respectively, so only \texttt{a}, \texttt{b} and \texttt{e} are in $\nd(r;\F)$. The gray area in Figure~\ref{fig:example} represents the union of $\F$-dominance regions $\dominanceRegion(\texttt{a};\F)\cup \dominanceRegion(\texttt{b};\F)\cup \dominanceRegion(\texttt{e};\F)$.
Finally, $\texttt{b}$ can never be top-$1$, since no linear combination of its scores can make it better than \texttt{a} or \texttt{e}; so, $\po(r;\F)$ consists of just \texttt{a} and \texttt{e}.
\end{example}

Among the ways to test $\F$-dominance when $\F$ is a set of functions under a set of linear constraints $\C$ on the weights, the most efficient one
consists in determining the $\F$-dominance region of the candidate $\F$-dominant tuple and checking whether the other tuple belongs to it. This 
 approach, described in Theorem~4.2 in~\cite{DBLP:journals/tods/CiacciaM20}, requires enumerating the vertices of the convex polytope defined by $\C$.
We will then consider the implementation denoted $\sveonef$~\cite{DBLP:journals/tods/CiacciaM20}, which uses vertex enumeration and also exploits pre-sorting of the dataset (as in SFS~\cite{DBLP:conf/icde/ChomickiGGL03}). Additionally, $\sveonef$ is a one-phase algorithm, in that it computes its result directly from $r$ instead of first computing $\sky(r)$ (which would be possible, since $\nd(r;\F)=\nd(\sky(r);\F)$) -- an approach that has been recognized to be more efficient if the constraints are tight enough.

Computing $\po$ can also be done 
through an LP problem.
Testing whether $t\in\po$ requires solving an expensive LP problem involving \emph{all} the tuples. An incremental approach is, however, possible: smaller LP problems with only a part of the tuples may be used to discard tuples. With this, we can solve LP problems of increasing sizes, starting from just two tuples, and then doubling such a number at each round, until all tuples are included.
We will adopt the implementation known as $\popi$\cite{DBLP:journals/tods/CiacciaM20}, which
 is incremental and works in two phases, i.e., computes its result from $\nd(r;\F)$ instead of $r$, observing that $\po(r;\F)=\po(\nd(r;\F);\F)$ and that for $\po$ this is much more efficient than a single phase.

\section{Parallel computation of flexible skylines}
\label{sec:parallel}

A general scheme for parallelizing the computation of flexible skylines is based on the idea that the dataset can be partitioned and each partition is processed independently and in parallel.
This produces a ``local'' result; the union of all local results still has some redundancies that can be eliminated by applying a last (sequential) round of removal.
This approach has been described for skylines, e.g., in~\cite{ciaccia2024optimizationstrategiesparallelcomputation}, and is correct, since, when $r=r_1 \cup \ldots \cup r_\partitions$, with $r_i\cap r_j=\emptyset$ for $i\neq j$, we have $\sky(r)=\sky(\sky(r_1)\cup\ldots\cup\sky(r_\partitions))$.
A similar principle applies to $\F$-skylines by replacing in the above formula $\sky(\cdot)$ with $\nd(\cdot;\F)$ or $\po(\cdot;\F)$.

\begin{algorithm}[t]
\scalebox{.95}
   {
    \begin{minipage}{1.33\textwidth}
	\begin{enumerate}[itemsep=0pt, parsep=0pt]
	    \item[Input:] \emph{relation $r$, functions $\F$ via constraints $\C$, number of partitions $\partitions$}
	    \item[Output:] \emph{a flexible skyline $\nd(r;\F)$ or $\po(r;\F)$}
		\item $ls := \emptyset$  // \textit{the local result}
		\item\label{line:partitioning-pattern} $(r_1,\ldots,r_\partitions, meta) := \texttt{Partition}(r,\partitions)$  // \textit{partitions and meta-information}
	    \item\label{line:parallel-for} \codeparallel\ \codeforeach\ $r_i$ in $r_1,\ldots,r_\partitions$ \codedo
	    \item\label{line:union-of-local-skylines} \quad $ls := ls \cup \texttt{ComputeLocalSet}(r_i,meta)$
	    \item\label{line:last-round-pattern} \codereturn\ $\nd(ls;\F)$ or $\po(ls;\F)$
	\end{enumerate}	    
	\end{minipage}
   }
	\caption{Algorithmic pattern for parallel skyline computation.}
	\label{alg:parallel-scheme}
\end{algorithm}

The algorithmic pattern in Algorithm~\ref{alg:parallel-scheme} does this in three phases: first $r$ is partitioned into $r_1,\ldots,r_\partitions$ (line~\ref{line:partitioning-pattern}), with possible meta-information to be used later; then, (local) results are computed independently (line~\ref{line:parallel-for}) and in $\codeparallel$, possibly using the meta-information;
finally, all the local results are merged (line~\ref{line:union-of-local-skylines}) and processed sequentially (line~\ref{line:last-round-pattern}).

\begin{figure}%
\centering
\subfloat[][{\grid}]
{\includegraphics[width=0.22\textwidth]{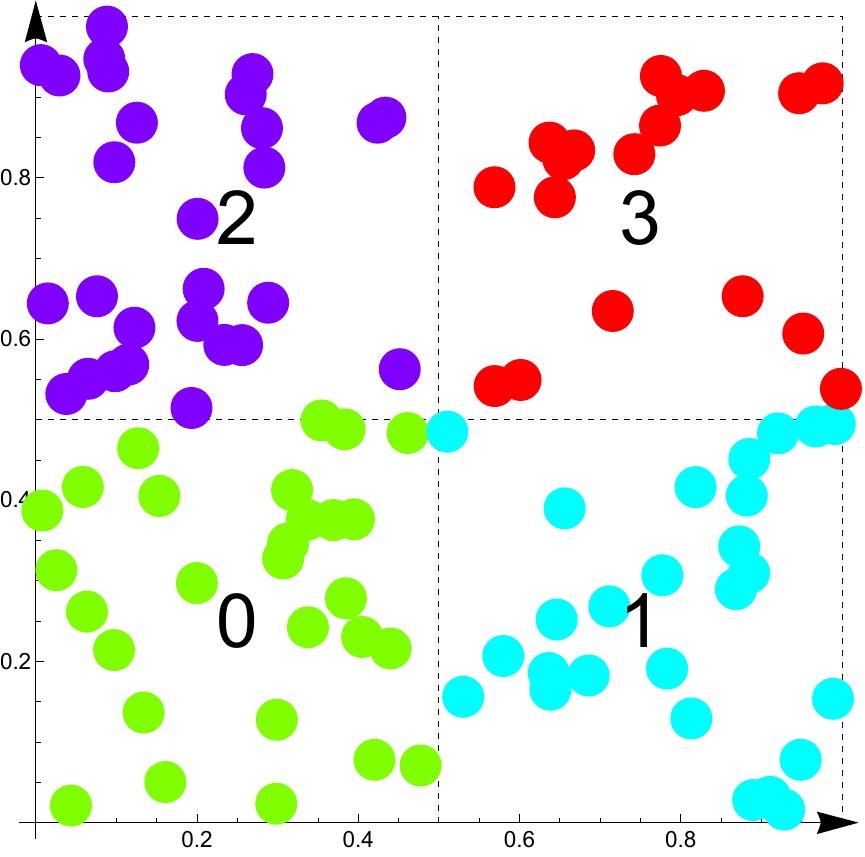}\label{fig:gridPartitioning}}%
\quad
\subfloat[][{\angular}]
{\includegraphics[width=0.22\textwidth]{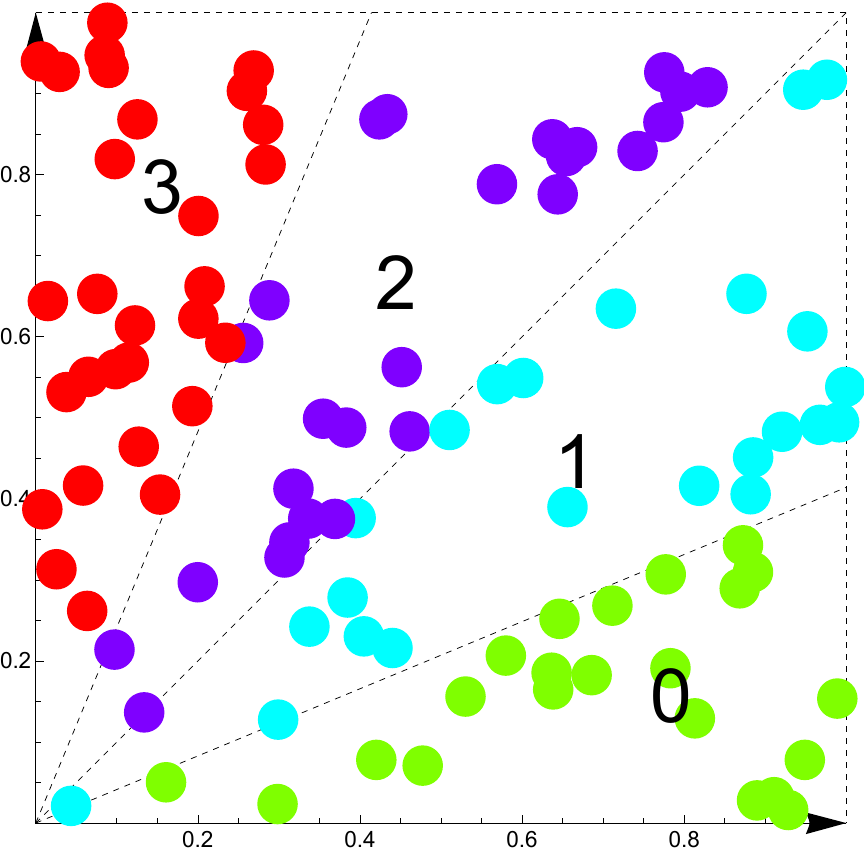}\label{fig:anglePartitioning}}%
\quad
\subfloat[][{\sliced}]
{\includegraphics[width=0.22\textwidth]{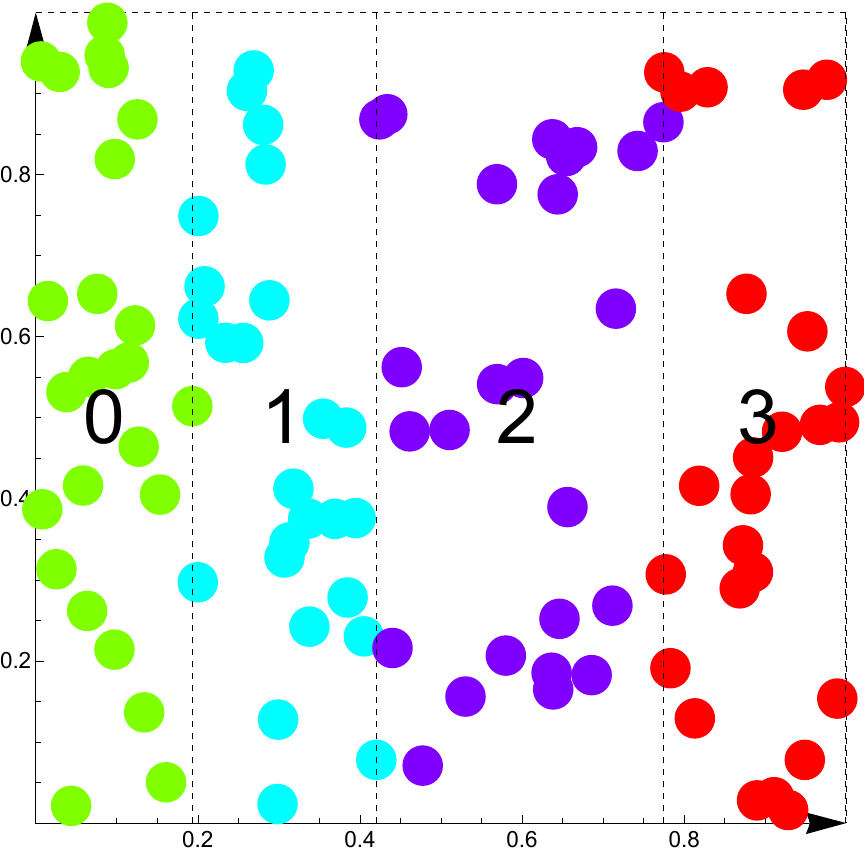}\label{fig:slicedPartitioning}}%
\caption{Partitioning Strategies.
 }\label{fig:precision-recall}%
\end{figure}

We shall adopt for flexible skylines the same partitioning strategies used in the literature for computing skylines, which are summarized next: Grid Partitioning, Angular Partitioning and Sliced Partitioning. We refrain from considering Random Partitioning~\cite{DBLP:conf/hpcs/Cosgaya-LozanoRZ07}, which simply randomly distributes the tuples across the various partitions, and is therefore too simple a baseline to receive further attention.

\subsection{Partitioning Strategies}\label{sec:partitioning}
\subsubsection{Grid Space Partitioning}

Grid Partitioning~\cite{DBLP:conf/edbt/MullesgaardPLZ14} (\grid) divides the space into a grid of equally sized cells.
Each dimension is divided into $\partitionsPerDim$ parts, resulting in a total of $\partitions=\partitionsPerDim^\dimensions$ partitions, where $\dimensions$ denotes the total number of dimensions.
We can also leverage dominance between grid cells to avoid processing certain partitions completely (\emph{Grid Filtering}).
In particular, let $c_i.\min$ and $c_i.\max$ represent the corners of cell $c_i$ with the lowest (best) or respectively highest (worst) values on all dimensions.
If $c_i.\max$ dominates $c_j.\min$ then all tuples in $c_i$ dominate all tuples in $c_j$, so if $c_i$ contains some tuple, $c_j$ can be disregarded.

With values in $[0,1]$, the partition for a tuple $t$ can be computed as follows:

\[
\partitions(t) = \sum_{i=1}^{\dimensions} \lfloor t[A_i]\cdot \partitionsPerDim \rfloor \cdot \partitionsPerDim^{i-1}
\]
\noindent where $A_i$ is the $i$-th attribute and $\partitionsPerDim$ are the slices per dimension.
Figure~\ref{fig:gridPartitioning} shows $\partitionsPerDim=2$ partitions per dimension, with a total of $4$ partitions (grid cells).

\subsubsection{Angle-based Space Partitioning}

Angle-based Partitioning~\cite{DBLP:conf/sigmod/VlachouDK08} (\angular) partitions the space based on angular coordinates, after converting Cartesian to hyper-spherical coordinates.
Unlike \grid, \angular does not support any kind of grid dominance.

The partition of tuple $t$
is computed based on hyper-spherical coordinates, including a radial coordinate $r$ and $\dimensions-1$ angular coordinates $\varphi_1, \ldots, \varphi_{\dimensions-1}$:

\begin{equation}\label{eq:index-angular}
\partitions(t)=\sum_{i=1}^{\dimensions-1}\left\lfloor \frac{2\varphi_i}{\pi}\partitionsPerDim \right\rfloor
\partitionsPerDim^{i-1}
\end{equation}
where $\partitionsPerDim$ is, again, the number of slices in which each (angular) dimension is divided, which amounts to grid partitioning on angular coordinates.
Figure~\ref{fig:anglePartitioning} shows \angular at work.

\subsubsection{Sliced Partitioning (One-dimensional Slicing)}

Sliced Partitioning (\sliced) sorts the dataset on one dimension.
The $i$-th tuple $t$ in the ordering is assigned to a partition in the following way:
\[
\partitions(t) = \left\lfloor \frac{(i-1)\cdot \partitions}{\size-1}\right\rfloor,
\]
where $\size$ is the number of tuples and $\partitions$ the number of partitions.
Figure~\ref{fig:slicedPartitioning} shows the effect of \sliced.

\subsection{Improvements}\label{sec:improvements}

A number of improvements are possible on top of the previously described algorithmic pattern.
The first one regards the choice of a selected set of tuples with a high potential for $\F$-dominating other tuples, which might be shared across all partitions from the start.
The second improvement consists in trying to eliminate the final sequential pass, which is typically the most time consuming.

\subsubsection{Representative Filtering}
A set of potentially ``strong'' tuples (the \emph{representatives}) from the entire dataset can be selected before entering the parallel phase. These are shared as meta-information across all partitions, so that any tuple dominated by a representative can be deleted with no further ado.
A simple selection criterion consists in selecting the first few tuples in each partition after sorting: by virtue of the topological sort property, they cannot be $\F$-dominated by subsequent tuples and therefore are likely to $\F$-dominate many of them.

\subsubsection{Elimination of sequential phase}
A way to eliminate the final sequential phase is,
instead of computing the global set sequentially from the union $U$ of the local sets, to re-partition $U$ into $u_1,\ldots,u_\partitions$ for a new parallel phase and also pass to each partition the entire $U$ as meta-information. With this, we can eliminate from each partition $u_i$ the (globally) $\F$-dominated tuples (i.e., those that are dominated by some tuple in $U$).
We call this scheme \noseq.

\section{Experiments}
\label{sec:experiments}

In this section, we measure the efficiency of the proposed algorithmic pattern for the computation of both $\nd$ and $\po$.
We measure our indicators of efficiency in different settings according to the operating parameters described in Table~\ref{tab:operating_parameters}. In particular, we target anticorrelated datasets, which are the most challenging in any skyline-like scenario. The set $\F$ of functions considered for computing flexible skylines is simply characterized by the constraint $w_1\geq w_2$, which prunes 50\% of the space of weights and can be applied to any dataset with at least $\dimensions\geq 2$ dimensions.

Our experiments are conducted on a computational infrastructure with Spark comprising virtual machines equipped with a total of 30 cores and 8GB of RAM.

\begin{table}[h]
    \centering
    \caption{Operating parameters for testing efficiency (defaults in bold).}
\scalebox{0.8}{
        \begin{tabular}{|l|l|}
            \hline
                Full name                           & Tested value \\
            \hline
                Dataset size ($\size$)                & 200K, 500K, \textbf{1M}, 2M, 5M, 10M\\
                \# of dimensions ($\dimensions$)    & 2, \textbf{4}, 6, 7 \\
                \# of partitions ($\partitions$)    & 10, 50, \textbf{100}, 150, 200, 300\\
                \# of cores ($\cores$)    & 5, 10, 20, \textbf{30} \\
            \hline
        \end{tabular}
        }
    \label{tab:operating_parameters}
\end{table}

\textbf{Varying the dataset size.}
Our first experiments test how efficiency varies as we vary the dataset size $\size$ and keep all other parameters set to a default value, i.e., $\dimensions=4$ dimensions,
$\partitions=100$ partitions, and $\cores=30$ cores.

\begin{figure}%
\centering
\subfloat[][{Total time}]
{\includegraphics[width=0.50\textwidth]{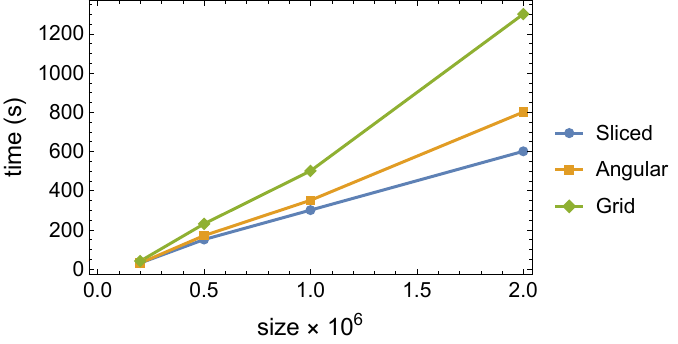}\label{fig:sveonef-size}}%
\subfloat[][{Parallel time}]
{\includegraphics[width=0.50\textwidth]{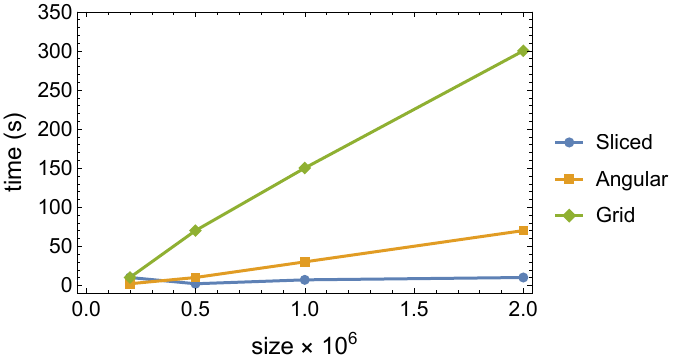}\label{fig:sveonef-size-parallel}}%
\caption{Performance of $\sveonef$ on an anticorrelated dataset with $\dimensions=4$ dimensions and varying sizes: total execution times (a); time for the parallel phase  (b).}\label{fig:sveonef-size-overall}.
\end{figure}

\begin{figure}%
\centering
\includegraphics[width=0.50\textwidth]{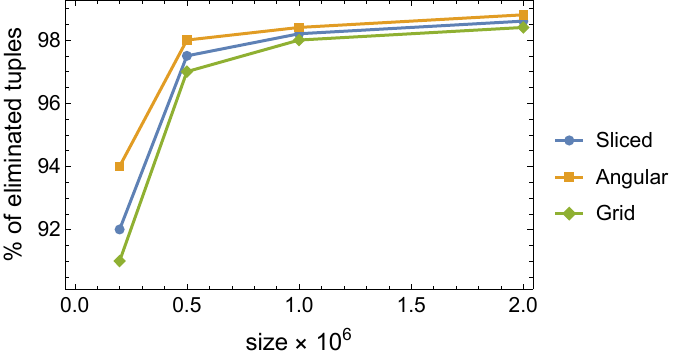}%
\caption{Percentage of removed tuples during the parallel phase by $\sveonef$ on an anticorrelated dataset with $\dimensions=4$ dimensions and varying sizes.}\label{fig:sveonef-size-perc}
\end{figure}

Figure~\ref{fig:sveonef-size} shows that, while of course execution times grow as the size grows, the \sliced strategy is much more efficient than \angular and, especially, \grid in exploiting parallelism during the computation of $\nd$ with $\sveonef$. In particular, Figure~\ref{fig:sveonef-size-parallel} shows the time spent during the parallel phase, highlighting how \grid goes astray and is inefficient in that many partitions are unbalanced, while the other two partitioning strategies are much more balanced in this respect.
Finally, Figure~\ref{fig:sveonef-size-perc} shows the percentage of tuples that are removed during the parallel phase, highlighting again that \grid is the least efficient choice, while \angular has the highest trimming power, although it is overall less efficient than \sliced because it pays an overhead due to the partial imbalance between partitions (which are perfectly balanced in case of \sliced).

Table~\ref{tab:sveonef-times} shows the breakdown of the execution times between the parallel and the sequential phase, along with the number of tuples surviving after the parallel phase, for a 4d dataset with 2M tuples, confirming the highest benefits of the \sliced strategy.

\begin{table}
\centering
\caption{Time of parallel and sequential phases with the $\sveonef$ Algorithm using a 4d
anticorrelated dataset with 2 million points and 16781 $\nd$ points.}
\begin{tabular}{|c|r|r|r|}
Partitioning & $|\bigcup \nd_i|$ & parallel phase ($s$) & sequential phase ($s$)\\
\hline
\grid & 39946 & 325.6 & 970.0\\
\hline
\angular & 33971 & 68.3 & 776.9\\
\hline
\sliced & 28051 & 14.8 & 679.3\\
\hline
\end{tabular}
	\label{tab:sveonef-times}
\end{table}

\begin{figure}%
\centering
\subfloat[][{Total time}]
{\includegraphics[width=0.50\textwidth]{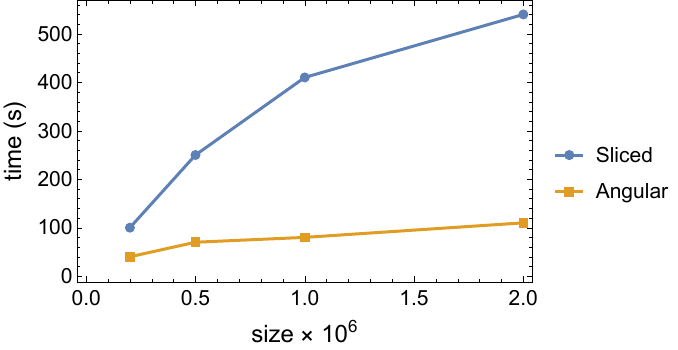}\label{fig:popi2-size}}%
\subfloat[][{Parallel time}]
{\includegraphics[width=0.50\textwidth]{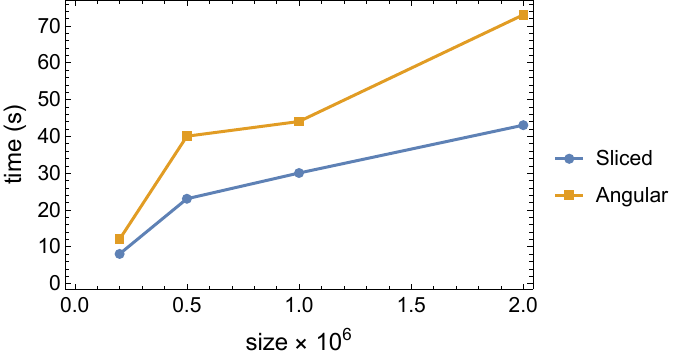}\label{fig:popi2-size-parallel}}%
\caption{Performance of $\popi$ on an anticorrelated dataset with $\dimensions=4$ dimensions and varying sizes: total execution times (a); time for the parallel phase (b).}\label{fig:popi2-size-overall}.
\end{figure}

\begin{figure}%
\centering
\includegraphics[width=0.50\textwidth]{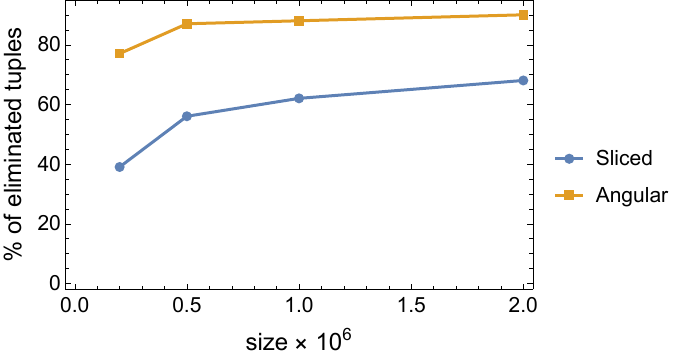}%
\caption{Percentage of removed tuples during the parallel phase by $\popi$ on an anticorrelated dataset with $\dimensions=4$ dimensions and varying sizes.}\label{fig:popi2-size-perc}.
\end{figure}

We perform a similar analysis with $\po$ using $\popi$.
We do not include results regarding \grid here, because, for solving LP problems of this size, we need partitions to be as balanced as possible, while \grid produces extremely unbalanced partitions (especially with anticorrelated data), with consequently very poor results.
Figure~\ref{fig:popi2-size} shows 
the times required to compute $\po$ from $\nd$ by using \popi (the time to compute $\nd$ is thus not included in the total).
We observe that, in this case, \sliced performs much more poorly than \angular, and this although \sliced still incurs lower execution times during the parallel phase, as shown in Figure~\ref{fig:popi2-size-parallel}. The reason is that the partitioning along one single dimension effected by \sliced is not particularly efficient at removing tuples during the parallel phase, as Figure~\ref{fig:popi2-size-perc} shows, with \angular removing as many as 85\% of the tuples with a 4d anticorrelated dataset with 2M tuples, while \sliced only attains 62\% in that case.
To be precise, $\po$ is a very small set compared to $\nd$ (with $|\nd|=16,781$ and $|\po|=150$ with the 4d dataset of 2M tuples), and this might be the source of ineffectiveness of sorting the dataset in one dimension as in \sliced.
Table~\ref{tab:popi-times} shows the execution times of the parallel and the sequential phase for a 4d dataset with 2 million tuples, highlighting that the \sliced strategy incurs a very high overhead in the sequential phase, due to a much worse ability to remove tuples during the parallel phase, with more than three times as many tuples remaining with the \sliced strategy than with the \angular strategy.

\begin{table}
\centering
\caption{Time of parallel and sequential phases with the $\popi$ Algorithm using a 4d
anticorrelated dataset with 2 million points and 16781 $\nd$ points and 150 $\po$ points.}
\begin{tabular}{|c|r|r|r|}
Partitioning & $|\bigcup \po_i|$ & parallel phase ($s$) & sequential phase ($s$)\\
\hline
\angular & 1544 & 72.4 & 48.9\\
\hline
\sliced & 5108 & 46.5 & 504.5\\
\hline
\end{tabular}
	\label{tab:popi-times}
\end{table}

\begin{figure}%
\centering
\subfloat[][{$\nd$ with $\sveonef$.}]
{\includegraphics[width=0.50\textwidth]{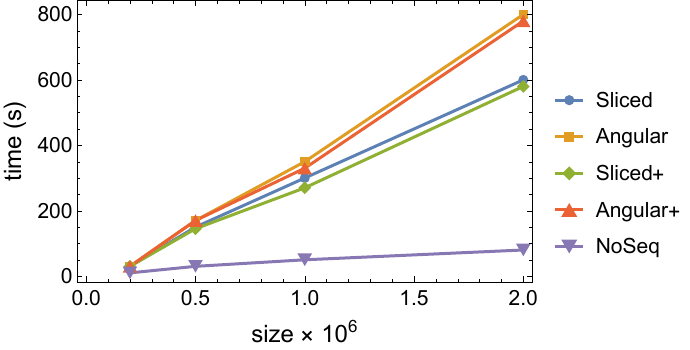}\label{fig:sveonef-size-improved}}%
\subfloat[][{$\po$ with $\popi$.}]
{\includegraphics[width=0.50\textwidth]{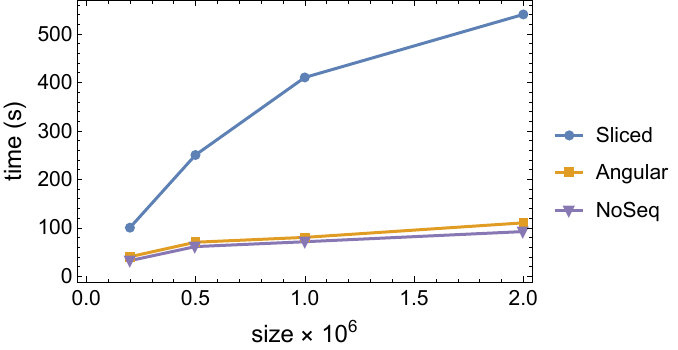}\label{fig:popi2-size-improved}}%
\caption{Performance of improved variants on an anticorrelated dataset with $\dimensions=4$ dimensions and varying sizes: total execution times for $\sveonef$ (a) and $\popi$ (b).}\label{fig:size-improved}.
\end{figure}

\medskip
\textbf{Varying size: improved strategies.}
Figure~\ref{fig:size-improved} shows the effect of applying the improvements discussed in Section~\ref{sec:improvements}.
Representative Filtering causes slight improvements in terms of execution times with both \sliced and \angular, as can be seen in Figure~\ref{fig:sveonef-size-improved} for the computation of $\nd$, where the improved versions are indicated as \slicedrep and \angularrep.
We also applied the \noseq scheme on top of the \sliced partitioning strategy (with representatives), which has determined significant improvements in the computation of $\nd$.
It turns out that \noseq largely outperforms the other strategies, with overall execution times that are more than 7 times smaller than the second best strategy (\slicedrep).
Grid Filtering, available for \grid partitioning, takes too long to return the local set and is overall disadvantageous in terms of duration compared to the basic version, so we will not consider it in the experiments.

As for the computation of $\po$ with $\popi$, Figure~\ref{fig:popi2-size-improved} reports the \noseq strategy along with \sliced and \angular (representatives prove to be ineffective for computing $\po$ and thus not reported). We observe that applying the \noseq scheme on top of \sliced makes it even more efficient than \angular, which was be the best option for this kind of problem.

\begin{figure}%
\centering
\includegraphics[width=0.50\textwidth]{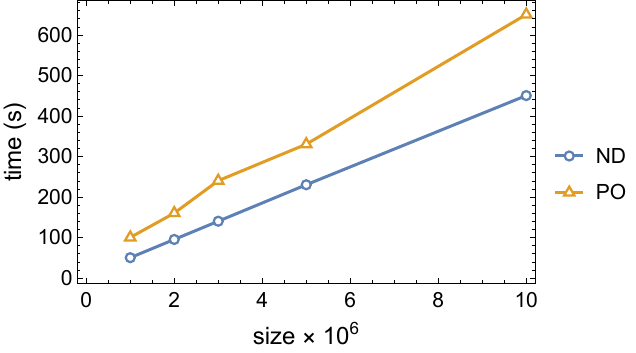}
\caption{Performance of the \noseq strategy on an anticorrelated dataset with $\dimensions=4$ dimensions and varying sizes: total execution times for $\sveonef$ and $\popi$.}\label{fig:size-improved-larger}.
\end{figure}

In order to complete our analysis as the size varies, we also tested the behavior of the \noseq strategy with even larger dataset sizes. Figure~\ref{fig:size-improved-larger} shows that, from 1M to 10M tuples, execution times incurred with \noseq follow an almost perfectly linear growth with the dataset size, both for $\nd$ and for $\po$, confirming scalability of the approach.

\begin{figure}%
\centering
\includegraphics[width=0.50\textwidth]{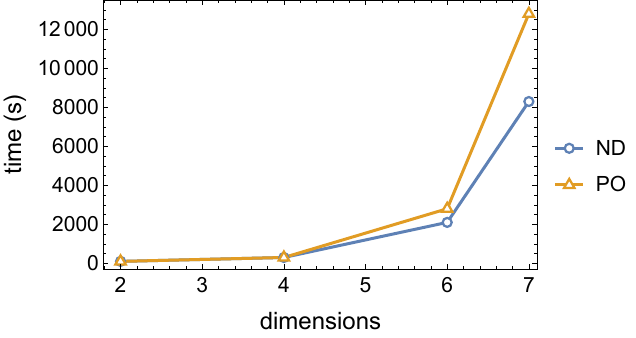}
\caption{Performance of the \noseq strategy on an anticorrelated dataset with size $\size=$1M and varying dimensions: total execution times for $\sveonef$ and $\popi$.}\label{fig:dim}.
\end{figure}

\medskip
\textbf{Varying dimensions.}
Increasing the number of dimensions $\dimensions$ results in a notable rise in execution time due to the significantly larger number of tuples in the output. As the number of dimensions increases, the probability of a tuple being $\F$-dominated decreases substantially. Figure~\ref{fig:dim} illustrates that, for anticorrelated datasets, \noseq is also affected by the ``curse of dimensionality'', with execution times increasing at a rate greater than linear as the number of dimensions grows.

\begin{figure}%
\centering
\includegraphics[width=0.50\textwidth]{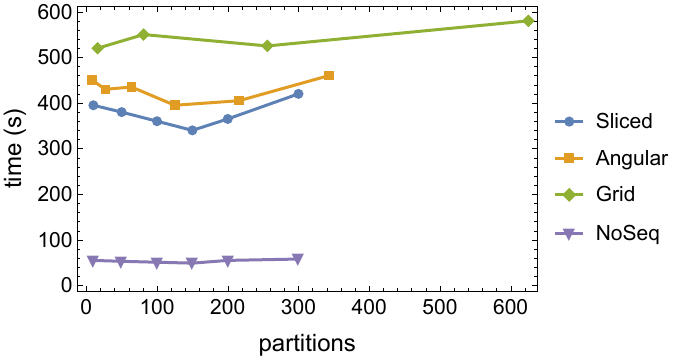}
\caption{Performance of various strategies for computing $\nd$ on a 4d anticorrelated dataset with size $\size=$1M and varying partitions.}\label{fig:partitions}.
\end{figure}

\medskip
\textbf{Varying number of partitions.}
The lowest execution times are reached when the number of partitions $\partitions$ is a small multiple of the number of available cores (30).
For larger values of $\partitions$, there is an increase in execution time due to synchronization overhead between nodes.
Figure~\ref{fig:partitions} shows this for several values of $\partitions$, with the extra observation that, while for \sliced (and \noseq, which is built on top of \slicedrep) all values of $\partitions$ are possible, \angular and \grid are constrained to have a number of partitions that equals $\partitionsPerDim^\dimensions$ and $\partitionsPerDim^{\dimensions-1}$, respectively, where $\partitionsPerDim$ is the number of slices per dimension.
The best value for $\partitions$ is around 150 (i.e., five times the number of available cores), after which performances start degrading. The \noseq strategy consistently outperforms the others for the computation of $\nd$.

\begin{figure}%
\centering
\subfloat[][{Computing $\nd$}]
{\includegraphics[width=0.50\textwidth]{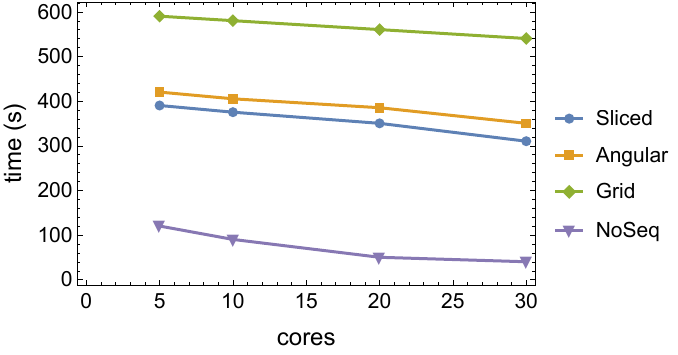}\label{fig:sveonef-cores}}%
\subfloat[][{Computing $\po$}]
{\includegraphics[width=0.50\textwidth]{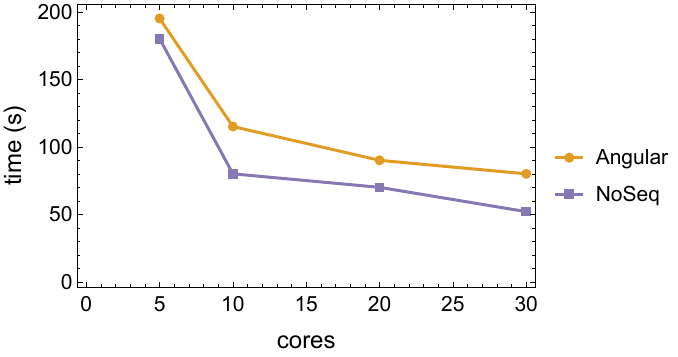}\label{fig:popi2-cores}}%
\caption{Effect of the number of cores $\cores$ on the execution time incurred by $\sveonef$ (a) and $\popi$ (b) on an anticorrelated dataset with $\dimensions=4$ dimensions and size $\size=$1M.}\label{fig:cores}.
\end{figure}

\medskip
\textbf{Varying number of cores.}
In our last experiment, we assess the effect of the number of cores $\cores$ on the execution time. Figure~\ref{fig:cores} shows benefits from the increased availability of cores, with the most visible effects near the beginning of the increase, when doubling the number of cores from 5 to 10. Adding further resources keeps improving execution times, but with more moderate benefits.
Overall, the \noseq strategy manages to improve its execution time by three times for computing $\nd$ and by $3.4$ times for computing $\po$ when moving from 5 to 30 cores.

\section{Related Work}
\label{sec:related}

There is an abundant literature that has studied and proposed sequential algorithms for the computation of the skyline, including Block-Nested Loop (BNL) and Divide-and-Conquer~\cite{DBLP:conf/icde/BorzsonyiKS01}, Sort Filter Skyline (SFS)~\cite{DBLP:conf/icde/ChomickiGGL03}, Bitmap~\cite{DBLP:conf/vldb/TanEO01}, Branch-and-Bound~\cite{DBLP:conf/sigmod/PapadiasTFS03,DBLP:journals/tods/PapadiasTFS05}, and Sort and Limit Skyline algorithm (SaLSa)~\cite{DBLP:conf/cikm/BartoliniCP06}.
The pre-sorting imposed by SFS is a suitable strategy for exploiting the topological sort property of dominance, which extends to $\F$-dominance, and has therefore been exploited in the sequential algorithms for computing flexible skylines~\cite{DBLP:journals/tods/CiacciaM20,DBLP:journals/pvldb/CiacciaM17,DBLP:conf/cikm/CiacciaM18,DBLP:conf/sebd/CiacciaM18,DBLP:conf/sisap/BedoCMO19,DBLP:conf/sebd/CiacciaM19,DBLP:conf/sigmod/MouratidisL021}), including $\sveonef$ and $\popi$, which we used here.

Systematic approaches for parallelizing the computation of skylines have been attempted for both \emph{vertical partitioning} (essentially following Fagin's middleware scenario for top-k queries~\cite{DBLP:conf/pods/Fagin98,DBLP:journals/tkde/TrimponiasBPY13,DBLP:journals/sigmod/ChomickiCM13}) and \emph{horizontal partitioning}~\cite{DBLP:journals/tkde/CuiCXLSX09}, which we considered in the present work. Horizontal partitioning strategies for computing the skyline have found their way to practical implementations with the advent of computation paradigms such as Map-Reduce and Spark, which allow seamless data distribution and processing through Mappers and Reducers, as demonstrated in~\cite{pindozzi,delorenzis,pinari,ciaccia2024optimizationstrategiesparallelcomputation}, together with optimization opportunities given by representative tuples and the elimination of the final sequential round, which was partially advocated in~\cite{DBLP:conf/edbt/MullesgaardPLZ14}.

While other works have considered flexible skylines in the context of vertical distribution~\cite{DBLP:conf/cikm/CiacciaM18,martinenghi2024computingnondominatedflexibleskyline}, the case of horizontal partitioning is, to the best of our knowledge, novel.

While classical skylines have no support for preferences, top-$k$ queries are the alternative (and, typically, preferred) ranking tool that not only supports preferences through scoring functions, but also controls the output size and accommodates various kinds of queries~\cite{DBLP:journals/csur/IlyasBS08,DBLP:journals/pvldb/MartinenghiT10,DBLP:journals/tkde/MartinenghiT12}. Many attempts are known that have tried to amalgamate top-$k$ queries and skylines so as to get the most out of these tools~\cite{DBLP:journals/pvldb/CiacciaM17,DBLP:conf/sigmod/MouratidisL021}.

Partitioning schemes similar to those that we have used in this paper have been successfully employed also for the computation of indicators of strength of skyline tuples, i.e., numeric values that can be adopted for endowing skyline tuples with further attributes to be used for ranking purposes.
These indicators include, e.g., the count of dominated tuples~\cite{DBLP:conf/sigmod/PapadiasTFS03}, the stability to perturbations in the scoring functions~\cite{DBLP:conf/sigmod/SolimanIMT11}, and the best possible rank of a tuple when using a linear scoring function~\cite{DBLP:journals/pvldb/MouratidisZP15}.
Several other indicators are introduced in~\cite{CM:PACMMOD2024}, some requiring to build the \emph{convex hull} of a dataset~\cite{DBLP:conf/pdcat/NakagawaMIN09,DBLP:conf/esa/WangYYD0S22,sym16121590}, or to compute complex volumes~\cite{DBLP:journals/csur/GuerreiroFP21,DBLP:journals/tcs/BringmannF12}; one of these indicators, called grid resistance, can be computed through parallel partitioning schemes similar to those discussed in this paper~\cite{parallelGridRes2024arXiv,C:MDPI2025},

Skylines, and more so flexible skylines, are common components of complex data preparation pipelines, typically followed by Machine Learning or Clustering algorithms~\cite{DBLP:conf/fqas/Masciari09,DBLP:journals/isci/MasciariMZ14,DBLP:conf/ismis/MasciariMPS20,DBLP:conf/ideas/MasciariMPS20} applied to data that might be collected by heterogeneous sources like RFID~\cite{DBLP:conf/ideas/FazzingaFMF09,DBLP:journals/tods/FazzingaFFM13}, pattern mining~\cite{DBLP:conf/ideas/MasciariGZ13}, crowdsourcing applications~\cite{DBLP:conf/socialcom/GalliFMTN12,DBLP:conf/www/BozzonCCFMT12,DBLP:conf/mmsys/LoniMGGMAMMVL13}, and streaming data~\cite{DBLP:journals/jiis/CostaMM14}.
In particular, the role of (flexible) skylines and top-$k$ queries is orthogonal to and typically integrated with that of query optimization and query explanation, which may rely on semantic properties of data expressible, e.g., through Query Containment (QC)~\cite{chandra1977optimal,shmueli87,DBLP:conf/er/CaliM08,CCM:JUCS2009} and integrity constraints~\cite{M:PHD2005,DBLP:conf/lopstr/ChristiansenM03,DBLP:conf/foiks/ChristiansenM04,DBLP:conf/fqas/Martinenghi04,DBLP:journals/access/SamarinA21,DBLP:conf/dexa/MartinenghiC05,DBLP:journals/aai/ChristiansenM00}.

While the preferences expressible with scoring functions and constraints on weights are of a quantitative nature, it would be interesting to know the extent to which qualitative preferences~\cite{DBLP:journals/jacm/CiacciaMT20} could also be integrated in the flexible skyline framework.

Other possible research avenues regard the stability of the results of a flexible skyline in the face of inconsistent or missing values~\cite{DBLP:conf/lpar/DeckerM06,DBLP:conf/dexaw/DeckerM06,DBLP:conf/dexaw/DeckerM07,DBLP:conf/ppdp/DeckerM08}, rather than uncertainty in the scoring function, and how this depends on the amount of such an inconsistency in the data~\cite{DBLP:journals/jiis/GrantH06,DBLP:conf/er/DeckerM09,DBLP:conf/ijcai/GrantH11,DBLP:conf/ecsqaru/GrantH13,DBLP:journals/ijar/GrantH17,DBLP:journals/ijar/GrantH23}.

\section{Conclusion}
\label{sec:conclusion}

In this paper, we tackled the problem of computing flexible skylines by exploiting common partitioning schemes that divide a dataset horizontally and send each partition to a different compute node for an initial round of computation of the result. The partial results, after this parallel phase, are collected and ultimately processed for a final, sequential round that eliminates all redundancies. While this consolidated scheme has been successfully used for computing the classical skyline operator, ours is the first attempt to apply it to the case of flexible skylines --- an extension of skylines that accommodates user preferences expressed through constraints on weights.

Our results show that the partitioning is generally beneficial, especially with schemes that divide the dataset into more balanced subsets, such as \angular and \sliced. In addition to optimization opportunities based on the notion of representative tuples, we also adopted another optimization, indicated as \noseq, that removes the final sequential round completely, which is typically the most expensive part of the entire computation. Indeed, \noseq proves to be the most robust of all tested algorithmic solutions.

The computation of the two flexible skyline operators, $\nd$ and $\po$, differs in the ability of the partitioning strategy to effectively reduce the set of candidates before the final round. We noticed that, while \sliced does its job egregiously in the case of $\nd$, the smaller sizes involved in the computation of potentially optimal tuples make \sliced a less suitable candidate than \angular for $\po$. In both cases, however, the \noseq strategy proves to be the best option.

Future research might try to address the impact of the constraints on weights on the effectiveness of the partitioning strategy.

\bibliographystyle{abbrv}

\begin{thebibliography}{10}

\bibitem{DBLP:conf/cikm/BartoliniCP06}
I.~Bartolini, P.~Ciaccia, and M.~Patella.
\newblock Salsa: computing the skyline without scanning the whole sky.
\newblock In P.~S. Yu, V.~J. Tsotras, E.~A. Fox, and B.~Liu, editors, {\em
  Proceedings of the 2006 {ACM} {CIKM} International Conference on Information
  and Knowledge Management, Arlington, Virginia, USA, November 6-11, 2006},
  pages 405--414. {ACM}, 2006.

\bibitem{DBLP:conf/sisap/BedoCMO19}
M.~V.~N. Bedo, P.~Ciaccia, D.~Martinenghi, and D.~de~Oliveira.
\newblock A k-skyband approach for feature selection.
\newblock In G.~Amato, C.~Gennaro, V.~Oria, and M.~Radovanovic, editors, {\em
  Similarity Search and Applications - 12th International Conference, {SISAP}
  2019, Newark, NJ, USA, October 2-4, 2019, Proceedings}, volume 11807 of {\em
  Lecture Notes in Computer Science}, pages 160--168. Springer, 2019.

\bibitem{DBLP:conf/icde/BorzsonyiKS01}
S.~B{\"{o}}rzs{\"{o}}nyi, D.~Kossmann, and K.~Stocker.
\newblock The skyline operator.
\newblock In {\em Proceedings of the 17th International Conference on Data
  Engineering, April 2-6, 2001, Heidelberg, Germany}, pages 421--430, 2001.

\bibitem{DBLP:conf/www/BozzonCCFMT12}
A.~Bozzon, I.~Catallo, E.~Ciceri, P.~Fraternali, D.~Martinenghi, and
  M.~Tagliasacchi.
\newblock A framework for crowdsourced multimedia processing and querying.
\newblock In {\em Proceedings of the First International Workshop on
  Crowdsourcing Web Search, Lyon, France, April 17, 2012}, pages 42--47, 2012.

\bibitem{DBLP:journals/tcs/BringmannF12}
K.~Bringmann and T.~Friedrich.
\newblock Approximating the least hypervolume contributor: Np-hard in general,
  but fast in practice.
\newblock {\em Theor. Comput. Sci.}, 425:104--116, 2012.

\bibitem{CCM:JUCS2009}
A.~Cal{\`{i}}, D.~Calvanese, and D.~Martinenghi.
\newblock {Dynamic Query Optimization under Access Limitations and
  Dependencies}.
\newblock {\em Journal of Universal Computer Science}, 15(21):33--62, 2009.
\newblock (SJR: Q2).

\bibitem{DBLP:conf/er/CaliM08}
A.~Cal{\`{\i}} and D.~Martinenghi.
\newblock Conjunctive query containment under access limitations.
\newblock In Q.~Li, S.~Spaccapietra, E.~S.~K. Yu, and A.~Oliv{\'{e}}, editors,
  {\em Conceptual Modeling - {ER} 2008, 27th International Conference on
  Conceptual Modeling, Barcelona, Spain, October 20-24, 2008. Proceedings},
  volume 5231 of {\em Lecture Notes in Computer Science}, pages 326--340.
  Springer, 2008.

\bibitem{chandra1977optimal}
A.~K. Chandra and P.~M. Merlin.
\newblock Optimal implementation of conjunctive queries in relational
  databases.
\newblock In {\em Proceedings of the 9th Annual ACM Symposium on Theory of
  Computing}, pages 77--90. ACM, 1977.

\bibitem{DBLP:journals/sigmod/ChomickiCM13}
J.~Chomicki, P.~Ciaccia, and N.~Meneghetti.
\newblock Skyline queries, front and back.
\newblock {\em {SIGMOD} Record}, 42(3):6--18, 2013.

\bibitem{DBLP:conf/icde/ChomickiGGL03}
J.~Chomicki, P.~Godfrey, J.~Gryz, and D.~Liang.
\newblock Skyline with presorting.
\newblock In U.~Dayal, K.~Ramamritham, and T.~M. Vijayaraman, editors, {\em
  Proceedings of the 19th International Conference on Data Engineering, March
  5-8, 2003, Bangalore, India}, pages 717--719. {IEEE} Computer Society, 2003.

\bibitem{DBLP:journals/aai/ChristiansenM00}
H.~Christiansen and D.~Martinenghi.
\newblock Symbolic constraints for meta-logic programming.
\newblock {\em Appl. Artif. Intell.}, 14(4):345--367, 2000.

\bibitem{DBLP:conf/lopstr/ChristiansenM03}
H.~Christiansen and D.~Martinenghi.
\newblock Simplification of database integrity constraints revisited: {A}
  transformational approach.
\newblock In M.~Bruynooghe, editor, {\em Logic Based Program Synthesis and
  Transformation, 13th International Symposium {LOPSTR} 2003, Uppsala, Sweden,
  August 25-27, 2003, Revised Selected Papers}, volume 3018 of {\em Lecture
  Notes in Computer Science}, pages 178--197. Springer, 2003.

\bibitem{DBLP:conf/foiks/ChristiansenM04}
H.~Christiansen and D.~Martinenghi.
\newblock Simplification of integrity constraints for data integration.
\newblock In D.~Seipel and J.~M.~T. Torres, editors, {\em Foundations of
  Information and Knowledge Systems, Third International Symposium, FoIKS 2004,
  Wilhelminenberg Castle, Austria, February 17-20, 2004, Proceedings}, volume
  2942 of {\em Lecture Notes in Computer Science}, pages 31--48. Springer,
  2004.

\bibitem{DBLP:journals/pvldb/CiacciaM17}
P.~Ciaccia and D.~Martinenghi.
\newblock Reconciling skyline and ranking queries.
\newblock {\em {PVLDB}}, 10(11):1454--1465, 2017.

\bibitem{DBLP:conf/sebd/CiacciaM18}
P.~Ciaccia and D.~Martinenghi.
\newblock Beyond skyline and ranking queries: Restricted skylines (extended
  abstract).
\newblock In S.~Bergamaschi, T.~D. Noia, and A.~Maurino, editors, {\em
  Proceedings of the 26th Italian Symposium on Advanced Database Systems,
  Castellaneta Marina (Taranto), Italy, June 24-27, 2018}, volume 2161 of {\em
  {CEUR} Workshop Proceedings}. CEUR-WS.org, 2018.

\bibitem{DBLP:conf/cikm/CiacciaM18}
P.~Ciaccia and D.~Martinenghi.
\newblock {FA} + {TA} {\textless} {FSA}: Flexible score aggregation.
\newblock In {\em Proceedings of the 27th {ACM} International Conference on
  Information and Knowledge Management, {CIKM} 2018, Torino, Italy, October
  22-26, 2018}, pages 57--66, 2018.

\bibitem{DBLP:conf/sebd/CiacciaM19}
P.~Ciaccia and D.~Martinenghi.
\newblock Flexible score aggregation (extended abstract).
\newblock In M.~Mecella, G.~Amato, and C.~Gennaro, editors, {\em Proceedings of
  the 27th Italian Symposium on Advanced Database Systems, Castiglione della
  Pescaia (Grosseto), Italy, June 16-19, 2019}, volume 2400 of {\em {CEUR}
  Workshop Proceedings}. CEUR-WS.org, 2019.

\bibitem{DBLP:journals/tods/CiacciaM20}
P.~Ciaccia and D.~Martinenghi.
\newblock Flexible skylines: Dominance for arbitrary sets of monotone
  functions.
\newblock {\em {ACM} Trans. Database Syst.}, 45(4):18:1--18:45, 2020.

\bibitem{CM:PACMMOD2024}
P.~Ciaccia and D.~Martinenghi.
\newblock {Directional Queries: Making Top-k Queries More Effective in
  Discovering Relevant Results}.
\newblock {\em Proc. {ACM} Manag. Data}, 2(6), 2024.
\newblock (GGS: A++).

\bibitem{ciaccia2024optimizationstrategiesparallelcomputation}
P.~Ciaccia and D.~Martinenghi.
\newblock Optimization strategies for parallel computation of skylines.
\newblock {\em CoRR}, arxiv:2411.14968, 2024.

\bibitem{DBLP:journals/jacm/CiacciaMT20}
P.~Ciaccia, D.~Martinenghi, and R.~Torlone.
\newblock Foundations of context-aware preference propagation.
\newblock {\em J. {ACM}}, 67(1):4:1--4:43, 2020.

\bibitem{DBLP:conf/hpcs/Cosgaya-LozanoRZ07}
A.~Cosgaya{-}Lozano, A.~Rau{-}Chaplin, and N.~Zeh.
\newblock Parallel computation of skyline queries.
\newblock In {\em 21st Annual International Symposium on High Performance
  Computing Systems and Applications {(HPCS} 2007), 13-16 May 2007, Saskatoon,
  Saskatchewan, Canada}, page~12. {IEEE} Computer Society, 2007.

\bibitem{DBLP:journals/jiis/CostaMM14}
G.~Costa, G.~Manco, and E.~Masciari.
\newblock Dealing with trajectory streams by clustering and mathematical
  transforms.
\newblock {\em J. Intell. Inf. Syst.}, 42(1):155--177, 2014.

\bibitem{DBLP:journals/tkde/CuiCXLSX09}
B.~Cui, L.~Chen, L.~Xu, H.~Lu, G.~Song, and Q.~Xu.
\newblock Efficient skyline computation in structured peer-to-peer systems.
\newblock {\em {IEEE} Trans. Knowl. Data Eng.}, 21(7):1059--1072, 2009.

\bibitem{delorenzis}
E.~{De Lorenzis}.
\newblock Computation of flexible skylines in a distributed environment, 2022.
\newblock {Master's Thesis}, Politecnico di Milano.

\bibitem{DBLP:conf/dexaw/DeckerM06}
H.~Decker and D.~Martinenghi.
\newblock Avenues to flexible data integrity checking.
\newblock In {\em 17th International Workshop on Database and Expert Systems
  Applications {(DEXA} 2006), 4-8 September 2006, Krakow, Poland}, pages
  425--429. {IEEE} Computer Society, 2006.

\bibitem{DBLP:conf/lpar/DeckerM06}
H.~Decker and D.~Martinenghi.
\newblock A relaxed approach to integrity and inconsistency in databases.
\newblock In M.~Hermann and A.~Voronkov, editors, {\em Logic for Programming,
  Artificial Intelligence, and Reasoning, 13th International Conference, {LPAR}
  2006, Phnom Penh, Cambodia, November 13-17, 2006, Proceedings}, volume 4246
  of {\em Lecture Notes in Computer Science}, pages 287--301. Springer, 2006.

\bibitem{DBLP:conf/dexaw/DeckerM07}
H.~Decker and D.~Martinenghi.
\newblock Getting rid of straitjackets for flexible integrity checking.
\newblock In {\em 18th International Workshop on Database and Expert Systems
  Applications {(DEXA} 2007), 3-7 September 2007, Regensburg, Germany}, pages
  360--364. {IEEE} Computer Society, 2007.

\bibitem{DBLP:conf/ppdp/DeckerM08}
H.~Decker and D.~Martinenghi.
\newblock Classifying integrity checking methods with regard to inconsistency
  tolerance.
\newblock In S.~Antoy and E.~Albert, editors, {\em Proceedings of the 10th
  International {ACM} {SIGPLAN} Conference on Principles and Practice of
  Declarative Programming, July 15-17, 2008, Valencia, Spain}, pages 195--204.
  {ACM}, 2008.

\bibitem{DBLP:conf/er/DeckerM09}
H.~Decker and D.~Martinenghi.
\newblock Modeling, measuring and monitoring the quality of information.
\newblock In C.~A. Heuser and G.~Pernul, editors, {\em Advances in Conceptual
  Modeling - Challenging Perspectives, {ER} 2009 Workshops CoMoL, ETheCoM,
  FP-UML, MOST-ONISW, QoIS, RIGiM, SeCoGIS, Gramado, Brazil, November 9-12,
  2009. Proceedings}, volume 5833 of {\em Lecture Notes in Computer Science},
  pages 212--221. Springer, 2009.

\bibitem{DBLP:conf/pods/Fagin98}
R.~Fagin.
\newblock Fuzzy queries in multimedia database systems.
\newblock In A.~O. Mendelzon and J.~Paredaens, editors, {\em Proceedings of the
  Seventeenth {ACM} {SIGACT-SIGMOD-SIGART} Symposium on Principles of Database
  Systems, June 1-3, 1998, Seattle, Washington, {USA}}, pages 1--10, 1998.

\bibitem{DBLP:journals/tods/FazzingaFFM13}
B.~Fazzinga, S.~Flesca, F.~Furfaro, and E.~Masciari.
\newblock Rfid-data compression for supporting aggregate queries.
\newblock {\em {ACM} Trans. Database Syst.}, 38(2):11, 2013.

\bibitem{DBLP:conf/ideas/FazzingaFMF09}
B.~Fazzinga, S.~Flesca, E.~Masciari, and F.~Furfaro.
\newblock Efficient and effective {RFID} data warehousing.
\newblock In B.~C. Desai, D.~Sacc{\`{a}}, and S.~Greco, editors, {\em
  International Database Engineering and Applications Symposium {(IDEAS} 2009),
  September 16-18, 2009, Cetraro, Calabria, Italy}, {ACM} International
  Conference Proceeding Series, pages 251--258, 2009.

\bibitem{DBLP:conf/socialcom/GalliFMTN12}
L.~Galli, P.~Fraternali, D.~Martinenghi, M.~Tagliasacchi, and J.~Novak.
\newblock A draw-and-guess game to segment images.
\newblock In {\em 2012 International Conference on Privacy, Security, Risk and
  Trust, PASSAT 2012, and 2012 International Conference on Social Computing,
  SocialCom 2012, Amsterdam, Netherlands, September 3-5, 2012}, pages 914--917,
  2012.

\bibitem{DBLP:journals/jiis/GrantH06}
J.~Grant and A.~Hunter.
\newblock Measuring inconsistency in knowledgebases.
\newblock {\em J. Intell. Inf. Syst.}, 27(2):159--184, 2006.

\bibitem{DBLP:conf/ijcai/GrantH11}
J.~Grant and A.~Hunter.
\newblock Measuring the good and the bad in inconsistent information.
\newblock In T.~Walsh, editor, {\em {IJCAI} 2011, Proceedings of the 22nd
  International Joint Conference on Artificial Intelligence, Barcelona,
  Catalonia, Spain, July 16-22, 2011}, pages 2632--2637. {IJCAI/AAAI}, 2011.

\bibitem{DBLP:conf/ecsqaru/GrantH13}
J.~Grant and A.~Hunter.
\newblock Distance-based measures of inconsistency.
\newblock In L.~C. van~der Gaag, editor, {\em Symbolic and Quantitative
  Approaches to Reasoning with Uncertainty - 12th European Conference,
  {ECSQARU} 2013, Utrecht, The Netherlands, July 8-10, 2013. Proceedings},
  volume 7958 of {\em Lecture Notes in Computer Science}, pages 230--241.
  Springer, 2013.

\bibitem{DBLP:journals/ijar/GrantH17}
J.~Grant and A.~Hunter.
\newblock Analysing inconsistent information using distance-based measures.
\newblock {\em Int. J. Approx. Reason.}, 89:3--26, 2017.

\bibitem{DBLP:journals/ijar/GrantH23}
J.~Grant and A.~Hunter.
\newblock Semantic inconsistency measures using 3-valued logics.
\newblock {\em Int. J. Approx. Reason.}, 156:38--60, 2023.

\bibitem{DBLP:journals/csur/GuerreiroFP21}
A.~P. Guerreiro, C.~M. Fonseca, and L.~Paquete.
\newblock The hypervolume indicator: Computational problems and algorithms.
\newblock {\em {ACM} Comput. Surv.}, 54(6):119:1--119:42, 2022.

\bibitem{DBLP:journals/csur/IlyasBS08}
I.~F. Ilyas, G.~Beskales, and M.~A. Soliman.
\newblock A survey of top-\emph{k} query processing techniques in relational
  database systems.
\newblock {\em {ACM} Comput. Surv.}, 40(4), 2008.

\bibitem{sym16121590}
H.~Kwon, S.~Oh, and J.-W. Baek.
\newblock Algorithmic efficiency in convex hull computation: Insights from 2d
  and 3d implementations.
\newblock {\em Symmetry}, 16(12), 2024.

\bibitem{DBLP:conf/mmsys/LoniMGGMAMMVL13}
B.~Loni, M.~Menendez, M.~Georgescu, L.~Galli, C.~Massari, I.~S. Alting{\"o}vde,
  D.~Martinenghi, M.~Melenhorst, R.~Vliegendhart, and M.~Larson.
\newblock Fashion-focused creative commons social dataset.
\newblock In {\em Multimedia Systems Conference 2013, MMSys '13, Oslo, Norway,
  February 27 - March 01, 2013}, pages 72--77, 2013.
\newblock (GGS: B-).

\bibitem{DBLP:conf/fqas/Martinenghi04}
D.~Martinenghi.
\newblock Simplification of integrity constraints with aggregates and
  arithmetic built-ins.
\newblock In H.~Christiansen, M.~Hacid, T.~Andreasen, and H.~L. Larsen,
  editors, {\em Flexible Query Answering Systems, 6th International Conference,
  {FQAS} 2004, Lyon, France, June 24-26, 2004, Proceedings}, volume 3055 of
  {\em Lecture Notes in Computer Science}, pages 348--361. Springer, 2004.

\bibitem{M:PHD2005}
D.~Martinenghi.
\newblock {\em Advanced Techniques for Efficient Data Integrity Checking}.
\newblock PhD thesis, Roskilde University, Dept. of Computer Science, Roskilde,
  Denmark, 2005.
\newblock Available in Datalogiske Skrifter, vol. 105, Roskilde University,
  Denmark.

\bibitem{martinenghi2024computingnondominatedflexibleskyline}
D.~Martinenghi.
\newblock Computing the non-dominated flexible skyline in vertically
  distributed datasets with no random access.
\newblock {\em CoRR}, arXiv:2412.15468, 2024.

\bibitem{parallelGridRes2024arXiv}
D.~Martinenghi.
\newblock Parallelizing the computation of robustness for measuring the
  strength of tuples.
\newblock {\em CoRR}, arXiv:2412.02274, 2024.

\bibitem{C:MDPI2025}
D.~Martinenghi.
\newblock {Parallelizing the Computation of Grid Resistance to Measure the
  Strength of Skyline Tuples}.
\newblock {\em Algorithms}, 18(1), 2025.

\bibitem{DBLP:conf/dexa/MartinenghiC05}
D.~Martinenghi and H.~Christiansen.
\newblock Transaction management with integrity checking.
\newblock In K.~V. Andersen, J.~K. Debenham, and R.~R. Wagner, editors, {\em
  Database and Expert Systems Applications, 16th International Conference,
  {DEXA} 2005, Copenhagen, Denmark, August 22-26, 2005, Proceedings}, volume
  3588 of {\em Lecture Notes in Computer Science}, pages 606--615. Springer,
  2005.

\bibitem{DBLP:journals/pvldb/MartinenghiT10}
D.~Martinenghi and M.~Tagliasacchi.
\newblock Proximity rank join.
\newblock {\em Proc. {VLDB} Endow.}, 3(1):352--363, 2010.

\bibitem{DBLP:journals/tkde/MartinenghiT12}
D.~Martinenghi and M.~Tagliasacchi.
\newblock Cost-aware rank join with random and sorted access.
\newblock {\em {IEEE} Trans. Knowl. Data Eng.}, 24(12):2143--2155, 2012.

\bibitem{DBLP:conf/fqas/Masciari09}
E.~Masciari.
\newblock Trajectory clustering via effective partitioning.
\newblock In T.~Andreasen, R.~R. Yager, H.~Bulskov, H.~Christiansen, and H.~L.
  Larsen, editors, {\em Flexible Query Answering Systems, 8th International
  Conference, {FQAS} 2009, Roskilde, Denmark, October 26-28, 2009.
  Proceedings}, volume 5822 of {\em Lecture Notes in Computer Science}, pages
  358--370, 2009.

\bibitem{DBLP:conf/ideas/MasciariGZ13}
E.~Masciari, S.~Gao, and C.~Zaniolo.
\newblock Sequential pattern mining from trajectory data.
\newblock In B.~C. Desai, J.~L. Larriba{-}Pey, and J.~Bernardino, editors, {\em
  17th International Database Engineering {\&} Applications Symposium, {IDEAS}
  '13, Barcelona, Spain - October 09 - 11, 2013}, pages 162--167. {ACM}, 2013.

\bibitem{DBLP:journals/isci/MasciariMZ14}
E.~Masciari, G.~M. Mazzeo, and C.~Zaniolo.
\newblock Analysing microarray expression data through effective clustering.
\newblock {\em Inf. Sci.}, 262:32--45, 2014.

\bibitem{DBLP:conf/ismis/MasciariMPS20}
E.~Masciari, V.~Moscato, A.~Picariello, and G.~Sperl{\`{\i}}.
\newblock A deep learning approach to fake news detection.
\newblock In D.~Helic, G.~Leitner, M.~Stettinger, A.~Felfernig, and Z.~W. Ras,
  editors, {\em Foundations of Intelligent Systems - 25th International
  Symposium, {ISMIS} 2020, Graz, Austria, September 23-25, 2020, Proceedings},
  volume 12117 of {\em Lecture Notes in Computer Science}, pages 113--122.
  Springer, 2020.

\bibitem{DBLP:conf/ideas/MasciariMPS20}
E.~Masciari, V.~Moscato, A.~Picariello, and G.~Sperl{\`{\i}}.
\newblock Detecting fake news by image analysis.
\newblock In B.~C. Desai and W.~Cho, editors, {\em {IDEAS} 2020: 24th
  International Database Engineering {\&} Applications Symposium, Seoul,
  Republic of Korea, August 12-14, 2020}, pages 27:1--27:5. {ACM}, 2020.

\bibitem{DBLP:conf/sigmod/MouratidisL021}
K.~Mouratidis, K.~Li, and B.~Tang.
\newblock Marrying top-k with skyline queries: Relaxing the preference input
  while producing output of controllable size.
\newblock In G.~Li, Z.~Li, S.~Idreos, and D.~Srivastava, editors, {\em {SIGMOD}
  '21: International Conference on Management of Data, Virtual Event, China,
  June 20-25, 2021}, pages 1317--1330. {ACM}, 2021.

\bibitem{DBLP:journals/pvldb/MouratidisZP15}
K.~Mouratidis, J.~Zhang, and H.~Pang.
\newblock Maximum rank query.
\newblock {\em {PVLDB}}, 8(12):1554--1565, 2015.

\bibitem{DBLP:conf/edbt/MullesgaardPLZ14}
K.~Mullesgaard, J.~L. Pederseny, H.~Lu, and Y.~Zhou.
\newblock Efficient skyline computation in mapreduce.
\newblock In S.~Amer{-}Yahia, V.~Christophides, A.~Kementsietsidis, M.~N.
  Garofalakis, S.~Idreos, and V.~Leroy, editors, {\em Proceedings of the 17th
  International Conference on Extending Database Technology, {EDBT} 2014,
  Athens, Greece, March 24-28, 2014}, pages 37--48. OpenProceedings.org, 2014.

\bibitem{DBLP:conf/pdcat/NakagawaMIN09}
M.~Nakagawa, D.~Man, Y.~Ito, and K.~Nakano.
\newblock A simple parallel convex hulls algorithm for sorted points and the
  performance evaluation on the multicore processors.
\newblock In {\em 2009 International Conference on Parallel and Distributed
  Computing, Applications and Technologies, {PDCAT} 2009, Higashi Hiroshima,
  Japan, 8-11 December 2009}, pages 506--511, 2009.

\bibitem{DBLP:conf/sigmod/PapadiasTFS03}
D.~Papadias, Y.~Tao, G.~Fu, and B.~Seeger.
\newblock An optimal and progressive algorithm for skyline queries.
\newblock In A.~Y. Halevy, Z.~G. Ives, and A.~Doan, editors, {\em Proceedings
  of the 2003 {ACM} {SIGMOD} International Conference on Management of Data,
  San Diego, California, USA, June 9-12, 2003}, pages 467--478. {ACM}, 2003.

\bibitem{DBLP:journals/tods/PapadiasTFS05}
D.~Papadias, Y.~Tao, G.~Fu, and B.~Seeger.
\newblock Progressive skyline computation in database systems.
\newblock {\em {TODS}}, 30(1):41--82, 2005.

\bibitem{pinari}
E.~Pinari.
\newblock Parallel implementations of the skyline query using pyspark, 2022.
\newblock {Master's Thesis}, Politecnico di Milano.

\bibitem{pindozzi}
A.~Pindozzi.
\newblock Scalable solutions for skyline computation using pyspark: Exploring
  parallel algorithms, 2023.
\newblock {Master's Thesis}, Politecnico di Milano.

\bibitem{DBLP:journals/access/SamarinA21}
S.~D. Samarin and M.~Amini.
\newblock Integrity checking for aggregate queries.
\newblock {\em {IEEE} Access}, 9:74068--74084, 2021.

\bibitem{shmueli87}
O.~Shmueli.
\newblock Decidability and expressiveness aspects of logic queries.
\newblock In {\em Proceedings of the sixth ACM SIGACT-SIGMOD-SIGART symposium
  on Principles of database systems}, pages 237--249. ACM Press, 1987.

\bibitem{DBLP:conf/sigmod/SolimanIMT11}
M.~A. Soliman, I.~F. Ilyas, D.~Martinenghi, and M.~Tagliasacchi.
\newblock Ranking with uncertain scoring functions: semantics and sensitivity
  measures.
\newblock In {\em Proceedings of the {ACM} {SIGMOD} International Conference on
  Management of Data, {SIGMOD} 2011, Athens, Greece, June 12-16, 2011}, pages
  805--816, 2011.

\bibitem{DBLP:conf/vldb/TanEO01}
K.~Tan, P.~Eng, and B.~C. Ooi.
\newblock Efficient progressive skyline computation.
\newblock In {\em {VLDB} 2001, Proceedings of 27th International Conference on
  Very Large Data Bases, September 11-14, 2001, Roma, Italy}, pages 301--310.
  Morgan Kaufmann, 2001.

\bibitem{DBLP:journals/tkde/TrimponiasBPY13}
G.~Trimponias, I.~Bartolini, D.~Papadias, and Y.~Yang.
\newblock Skyline processing on distributed vertical decompositions.
\newblock {\em {IEEE} Trans. Knowl. Data Eng.}, 25(4):850--862, 2013.

\bibitem{DBLP:conf/sigmod/VlachouDK08}
A.~Vlachou, C.~Doulkeridis, and Y.~Kotidis.
\newblock Angle-based space partitioning for efficient parallel skyline
  computation.
\newblock In J.~T. Wang, editor, {\em Proceedings of the {ACM} {SIGMOD}
  International Conference on Management of Data, {SIGMOD} 2008, Vancouver, BC,
  Canada, June 10-12, 2008}, pages 227--238. {ACM}, 2008.

\bibitem{DBLP:conf/esa/WangYYD0S22}
Y.~Wang, R.~Yesantharao, S.~Yu, L.~Dhulipala, Y.~Gu, and J.~Shun.
\newblock Pargeo: {A} library for parallel computational geometry.
\newblock In S.~Chechik, G.~Navarro, E.~Rotenberg, and G.~Herman, editors, {\em
  30th Annual European Symposium on Algorithms, {ESA} 2022, September 5-9,
  2022, Berlin/Potsdam, Germany}, volume 244 of {\em LIPIcs}, pages
  88:1--88:19, 2022.

\end{thebibliography}

\end{document}